\documentclass[12pt]{article}
\usepackage{amsmath,amssymb}
\usepackage{graphicx,color,xcolor}
\usepackage{framed,multirow,cancel}
\usepackage{hyperref,cleveref}
\usepackage{ulem}
\colorlet{framecolor}{black}
\colorlet{shadecolor}{lightgray}
 \allowdisplaybreaks
\newcommand{\be}{\begin{equation}}
\newcommand{\ee}{\end{equation}}
\newcommand{\bea}{\begin{eqnarray}}
\newcommand{\eea}{\end{eqnarray}}

\newcommand{\nn}{\nonumber}

\newcommand{\wtd}{\widetilde}

\long\def\symbolfootnote[#1]#2{\begingroup%
	\def\thefootnote{\fnsymbol{footnote}}\footnote[#1]{#2}\endgroup}

\newcommand{\hust}{\it MoE Key Laboratory of Fundamental Quantities Measurement and School of Physics,\\
	Huazhong University of Science and Technology,\\
	Luoyu Lu 1037, Wuhan 430074, China}

\newcommand{\sysu}{\it TianQin Research Center for Gravitational Physics and School of Physics and Astronomy\\
    Sun Yat-sen University (Zhuhai Campus), Zhuhai 519082, China}

\begin{document}

\thispagestyle{empty}
\begin{center}

~\vspace{20pt}

{\Large\bf Extended symmetries at black hole horizons in generic dimensions}

\vspace{25pt}

\vspace{25pt}

	Changfu Shi\symbolfootnote[1]{Email:~\sf cfshi@hust.edu.cn}
    and Jianwei Mei\symbolfootnote[2]{Email:~\sf meijw@sysu.edu.cn}

\vspace{10pt}

$^\ast$\hust

\vspace{10pt}

$^\dagger$\sysu

\vspace{2cm}

\centerline{\bf Abstract}
\end{center}

Recently it has been shown that there is asymptotic BMS-like symmetry associated with the near-horizon geometry of black holes in three and four dimensions. In this paper, we show that the presence of such BMS-like symmetry is a ubiquitous feature for black holes in generic dimensions. For black holes in $D$ dimensions, the symmetry contains 2 supertranslations and $D-2$ generalized superrotations. The superrotations are found to generate a generalized Witt-like algebra that was previously noticed in a rather different construction. In the case of stationary and axisymmetric black holes, we calculate the surface charges and show that the zero-mode charges are intimately related to the entropy and angular momenta of the black hole.

\newpage
\section{INTRODUCTION}
In the past few years, important connections have been made between BMS symmetries and soft theorems\cite{SM1,SM2,SM3,SM4}. The BMS symmetries \cite{BMS1,BMS2} transform the Minkowski vacuum into other physically inequivalent vacua, which differ from each other by the creation or annihilation of soft gravitons. All such vacua have zero energy but with different angular momenta. Hawking, Perry and Strominger \cite{SFBH} suggested that these inequivalent vacua might play a pivotal role in solving the information paradox. They argued that the ``losing" information could be stored in the supertranslation and superrotation charges on the horizon, so called ``soft hair," associated with the shifts of the horizon which are caused by the ingoing particles \cite{IPBH}. An alternative interpretation of the symmetries can be found in \cite{SOBHSH}.

BMS-like symmetries are previously only constructed at the null infinity of asymptotically flat spacetimes. In \cite{SNIDTH}, Booth constructed the near horizon metrics for $D$ dimensional black holes, by starting with a $D-2$ dimensional hypersurface. These near horizon metrics are applicable to any $D$ dimensional black holes. Using these metrics and with appropriate boundary conditions, Donnay {\it et al} demonstrated the existence of BMS-like supertranslation and superrotation symmetries near the horizons of black holes in three and four dimensions \cite{ESBH,STSR}. Using BTZ \cite{BTZ1,BTZ2} and Kerr black hole as explicit examples, they found that the nonvanishing zero-mode charges are related to the black hole entropy and angular momentum. The importance of this work was soon noticed in several other works, e.g. \cite{HSDBHS,BMSNIH,NHGWCS,MGBC,SHOSH,ASNI}.

In this paper, we would like to investigate the same problem in generic dimensions. This is technically made possible by the general near horizon metric available from \cite{SNIDTH} and the existing strategy to calculate charges using near horizon data for stationary and axisymmetric black holes in generic dimensions \cite{JWM2}. The main motivation for the effort is that, due to the possibility for more independent rotations in higher dimensions, the near horizon BMS-like symmetries as found in \cite{ESBH,STSR} will very likely contain more generalized superrotations, and superrotation is known to play a significant role in the recent discussion on ``missing" information for black holes \cite{SFBH,SCSHBH,CBMS,SBHPC,VGF,CSFSCSP}. By moving to generic dimensions and making the number of generalized superrotations a variable, it could be easier to test possible scenarios on how exactly the BMS-like symmetries can contribute to the quantity of accountable ``missing" information. We find that there also exist infinite-dimensional symmetries near the horizons of black holes in generic dimensions. In $D$ dimensions, these symmetries contain 2 supertranslations and $D-2$ generalized superrotations. The superrotations are found to generate a new Witt-like algebra that was only recently noticed but in a rather different construction \cite{JWM3}.

The paper is organized as follows. In Sec. 2, we discuss the extended symmetries of general $D$-dimensional black holes, following the same logic steps of \cite{STSR}. In Sec.3, we apply the result to the case of stationary black holes and calculate the charges. Sec.4 contains a brief summary.

\section{SUPERTRANSLATION AND SUPERROTATION NEAR D-DIMENSIONAL BLACK HOLE HORIZONS}

\subsection{Boundary conditions}
In \cite{SNIDTH}, Booth constructed the near horizon metrics for $D$ dimensional black holes, by starting with a $D-2$ dimensional hypersurface. According to \cite{SNIDTH}, consider a D-dimensional black hole, whose near horizon geometry can be characterized by Gaussian null coordinates
\begin{equation}
\label{metric1}
ds^2=g_{vv}dv^2 +2dv d\rho+2g_{vA}dv dx^A +g_{AB}dx^A dx^B
\end{equation}
where  \(\upsilon\) is the advanced time coordinate, \( \rho\geq0\) describes the radial distance to the horizon, \(x^A\)\((A=1,\dots,D-2)\) are the angular coordinates.

Similar with \cite{ESBH,STSR}, the boundary conditions that chosen in this paper are
\begin{align}
\label{bd}
g_{vv}&=-2\kappa\rho +\mathcal{O}(\rho^2)\notag\\
g_{vA}&=\rho \theta_A +\mathcal{O}(\rho^2)\notag\\
g_{AB}&=\Omega_{AB}+\rho \lambda_{AB}+\mathcal{O}(\rho^2)
\end{align}
where \(\kappa,\theta_A,\lambda_{AB}\) and \(\Omega_{AB}\) are functions of \(\upsilon\) and \(x^A\). The indices \{A\} are risen and lowered by the metric on angular coordinates \(\Omega_{AB}\), while A, B and C...stand for arbitrary angular coordinates. \(\mathcal{O}(\rho^2)\) items stand for those functions vanish equally or faster than \(\rho^2\) at small \(\rho\). Some metric elements which are not mentioned in Eq.(\ref{metric1}) are regarded as the same order or higher order than \(\mathcal{O}(\rho^2)\).

In other words, the metrics behave in the near horizon region as
\begin{equation}
\label{mabc}
ds^2 = -2\kappa\rho d\upsilon^2 + 2 d\rho d\upsilon + 2 \rho \theta_A d\upsilon dx^A + (\Omega_{AB} + \rho \lambda_{AB}) dx^A dx^B + \Delta g_{ij}(\rho^2) dx^i dx^j .
\end{equation}
It is always possible to get a coordinate system to admits the form Eq.(\ref{mabc}) which could describe the black hole horizon.

\subsection{Asymptotic Killing vectors and symmetries}
In the near horizon region, the asymptotic Killing vectors can be expanded as
\begin{equation}
\begin{split}
\xi^\upsilon &= f_{11} + \rho f_{12} + \rho^2 f_{13} + \mathcal{O}(\rho^3),\\
\xi^\rho &=f_{21} + \rho f_{22} + \rho^2 f_{23}+\mathcal{O}(\rho^3),\\
\xi^A &=f_{31}^A + \rho f_{32}^A + \rho^2 f_{33}^A+\mathcal{O}(\rho^3).
\end{split}
\end{equation}
All the functions in above expressions only depend on \(\upsilon\) and \(x^A\). As the asymptotic Killing vectors must preserve the boundary conditions Eq.(\ref{bd}),  \({\mathcal{L}}_\xi g_{ij} = \delta_\xi g_{ij}\), which means the variations of the metric along the vectors will keep the form of
\begin{align}
&\mathcal{L}_\xi g_{\upsilon\upsilon} = - 2 \rho \delta_\zeta \kappa + \mathcal{O}(\rho^2), \label{LD1}\\
&\mathcal{L}_\xi g_{\upsilon A} = \rho \delta_\zeta \theta_A + \mathcal{O}(\rho^2) ,\label{LD2}\\
&\mathcal{L}_\xi g_{AB} = \delta_\xi \Omega_{AB} + \rho \delta_\xi \lambda_{AB} + \mathcal{O}(\rho^2), \label{LD3}\\
&\mathcal{L}_\xi g_{\rho\rho} = \mathcal{O}(\rho^2) ,~~~ \mathcal{L}_\xi g_{\upsilon\rho} = \mathcal{O}(\rho^2) ,~~~ \mathcal{L}_\xi g_{\rho A} = \mathcal{O}(\rho^2) .\label{LD4}
\end{align}

By calculating the Lie derivative of Eq.(\ref{mabc}) along \(\xi\), one can get the following equations from  Eq.(\ref{LD4})
\begin{align}
f_{12} &= 0, f_{13} = 0\\
f_{22} &= -  \partial_\upsilon f_{11} ,  f_{23}=\dfrac{1}{2} \Omega^{AB} \theta_A \partial_B f\\
f_{32} ^A &= - \Omega^{AB} \partial_B f_{11} , f_{33}^A=\frac{1}{2} \Omega^{AB} \Omega^{CD} \lambda_{BD} \partial_C f.
\end{align}
And Eq.(\ref{LD1}) leads to
\begin{align}
&\delta_\xi \kappa = f_{31}^A \partial_A \kappa + \partial_\upsilon ^2 f_{11} - \theta_A \partial_\upsilon f_{31}^A + \kappa \partial_\upsilon f_{11} + f_{11} \partial_\upsilon \kappa\\
&\partial_\upsilon f_{21} = \kappa f_{21}. \label{SF1}
\end{align}
At the same time, Eq.(\ref{LD2}) leads to
\begin{align}
\partial_A f_{21} =& - \theta_A f_{21} - \Omega_{AC}\partial_\upsilon f_{31} ^C  \label{SF2}\\
\delta_\xi \theta_A =& -2\kappa \partial_A f_{11} + \theta_B \partial_A f_{31}^B + f_{31}^B \partial_B \theta_A + f_{11} \partial_\upsilon \theta_A - 2\partial_\upsilon \partial_A f_{11} \\ \notag
&+ \Omega^{BC} \partial_\upsilon \Omega_{AB} \partial_C f_{11} + \lambda_{AB} \partial_\upsilon f_{31}^B.
\end{align}

Assuming that the leading items of the Killing vector does not depend on the fields. Eq.(\ref{SF2}) implies
\begin{align}
f_{21} = 0 ,~~~~ \partial_\upsilon f_{31} ^A =0.
\end{align}

Therefore, the form of asymptotic Killing vectors that meet boundary conditions can be given as
\begin{equation}
\label{akv}
\begin{split}
&\xi^\upsilon = f(\upsilon,x^A) +\mathcal{O}(\rho^3) ,\\
&\xi^\rho = -\rho \partial_ \upsilon f + \dfrac{1}{2} \rho^2 \Omega^{AB} \theta_A \partial_B f + \mathcal{O}(\rho^3),\\
&\xi^A = Y^A(x^E) - \rho \Omega^{AC} \partial_C f +\frac{1}{2} \rho^2 \Omega^{AB} \Omega^{CD} \lambda_{BD} \partial_C f + \mathcal{O}(\rho^3).
\end{split}
\end{equation}
For a given metric, its asymptotic Killing vectors can be regarded as just functions of \(f\) and \(Y^A\). Note that, each angular coordinate corresponds to a \(Y^A\), it means that there exist \(D-2\) pieces of \(Y^A\). Particularly, there are only two pieces of \(Y^A\) in four dimensional cases.

Then according to Eq.(\ref{LD1})--Eq.(\ref{LD3}), the corresponding variation of the fields occur in the metric
\begin{align}
& \delta_\xi \kappa = Y^A \partial_A \kappa + \partial_\upsilon ^2 f + \partial_\upsilon (\kappa f), \label{cvf1}\\
& \delta_\xi \theta_A = \mathcal{L}_Y \theta_A + f \partial_\upsilon \theta_A -2\kappa \partial_A f -2 \partial_A \partial_\upsilon f - \Omega_{AB} \partial_\upsilon \Omega^{BC} \partial_C f , \label{cvf2}\\
& \delta_\xi \Omega_{AB} = \mathcal{L}_Y \Omega_{AB} + f \partial_\upsilon \Omega_{AB} , \label{cvf3}\\
& \delta_\xi \lambda_{AB} = \mathcal{L}_Y \lambda_{AB} + \theta_A \partial_B f +\theta_B \partial_A f +f \partial_\upsilon \lambda_{AB} - \lambda_{AB} \partial_\upsilon f - 2 \nabla_A \nabla_B f  ,\label{cvf4}
\end{align}
where \(\nabla_A\) denotes the covariant derivative corresponding to \(\Omega_{AB}\) and \(\mathcal{L}_Y\) stands for the Lie derivative along \(Y^A\). 

It is easy to see that the asymptotic Killing vectors are locally  depend on fields defined in the metric, and the algebra generated by Lie brackets does not close. By taking a modified version of Lie brackets mentioned in \cite{ABCC,BMSSA}
\begin{align}
[\xi_1 , \xi_2] \equiv \mathcal{L}_{\xi_1}\xi_2 - \delta_{\xi_1} \xi_2 + \delta_{\xi_2} \xi_1 .
\end{align}
The algebra of asymptotic Killing vectors is given by
\begin{align}
\label{caeg}
[\xi(f_1 , Y^A_1) , \xi(f_2 , Y^A_2)] = \xi(f_{12} , Y^A_{12}),
\end{align}
then the corresponding algebra reads:
\begin{align}
& f_{12} = f_1 \partial_\upsilon f_2 - f_2 \partial_\upsilon f_1 + Y_1^A \partial_A f_2 -Y_2^A \partial_A f_1 \label{cfe1}\\
& Y^A_{12} = Y^B_1 \partial_B Y^A_2 - Y^B_2 \partial_B Y^A_1 \label{cfe2}.
\end{align}

\subsection{Charges}
The variation of surface charges for arbitrary dimensional black holes caused by asymptotic Killing vectors can be calculated in covariant approach \cite{CTAS,SCAGT}
\begin{equation}
\begin{split}
\label{vsce}
\cancel{\delta} Q_\xi [g;h]=\dfrac{1}{16\pi G}\int (d^{n-k} x)_{\mu\nu}\sqrt{-g}&[\xi^\nu \nabla^\mu h-\xi ^\nu \nabla_\sigma h^{\mu\sigma}+\xi_\sigma \nabla^\nu h^{\mu\sigma}\\
&+\dfrac{1}{2}  h \nabla^\nu \xi^\mu +\dfrac{1}{2} h^{\nu\sigma}(\nabla^\mu \xi_\sigma-\nabla_\sigma\xi^\mu)],\\
\end{split}
\end{equation}
where
\begin{align}
(d^{n-k} x)_{\mu\nu}=\dfrac{1}{k!(n-k)!}\varepsilon_{\nu\mu \alpha_1 \alpha_1 ...\alpha_{n-2}}dx^{\alpha_1}\wedge dx^{\alpha_2}\wedge...dx^{\alpha_{n-2}} .
\end{align}
Here \(\xi\) is an specific asymptotic Killing vector which is given by specific  \(f\) and \(Y^A\), and \(h_{\mu\nu}=\delta g_{\mu\nu}\) is the variation of the metric under this \(\xi\), \(h\) is the trace of \(h_{\mu\nu}\). The symbol \(\cancel{\delta}\) indicates that this expression may not be integrated out.

Then the charge on the horizon can be calculated:
\begin{equation}
\begin{split}
\label{vscd}
\cancel{\delta} Q_{[f;Y^A]}=&\dfrac{1}{16\pi G}\int (d^{n-2}x)_{\upsilon\rho}2[2f\kappa \delta(\sqrt{\det\Omega})+2\partial_\upsilon f \delta(\sqrt{\det\Omega})-2f\partial_\upsilon \delta(\sqrt{\det\Omega})\\
&-Y^A\delta(\theta_A\sqrt{\det\Omega})+\dfrac{1}{2}f\sqrt{\det\Omega}(\Omega^{AB}\Omega^{CD}-\Omega^{AC}\Omega^{BD})\partial_\upsilon \Omega_{CD}\delta \Omega_{AB}]\\
\end{split}
\end{equation}
This expression cannot be integrated out, as \(\kappa\) can vary under \(\xi\), i.e. \(\delta\kappa\neq 0\), and it cannot be expressed by any functions of \(\Omega\). This problem can be avoided for the cases with fixed \(\kappa\).  The last item cannot be integrated out either because it involves both \(\Omega_{AB}\) and its derivative. But when discussing the cases of stationary black holes, the last item  will vanish also.

\subsection{Algebra}
Now consider the case of isolated horizon with fixed temperature. Assuming that \(\kappa\) is a constant, Eq.(\ref{cvf1})-Eq.(\ref{cvf4}) can be simplified with fixed \(\kappa\), then Eq.(\ref{cvf1}) transforms into
\begin{equation}
\kappa \partial_\upsilon f + \partial_\upsilon^2 f = 0.
\end{equation}
The form of its solutions need to be
\begin{align}
f(\upsilon , x^A) = T(x^A) + e^{-\kappa \upsilon} X(x^A)
\end{align}
Then the relation in Eq.(\ref{cfe1}) and Eq.(\ref{cfe2}) will be
\begin{align}
& T_{12} = Y^A_1 \partial_A T_2 -Y^A_2 \partial_A T_1 ,\\
& X_{12} = Y^A_1 \partial_A X_2 -Y^A_2 \partial_A X_1 - \kappa (T_1X_2 - T_2X_1) ,\\
& Y^A_{12} = Y^B_1 \partial_B Y^A_2 - Y^B_2 \partial_B Y^A_1 .
\end{align}

The asymptotic Killing vectors can be expressed as \(\xi = \xi (T , X , Y^A,Y^B ,\dots)\), and, without loss of generality, the functions can be expanded in Laurent modes,
\begin{equation}
\begin{split}
\label{ELM}
& T_{(m,n,\dots,p,\dots ,l)} = \xi ((x^1)^m (x^2)^n\dots (x^a)^p\dots (x^{D-2})^l , 0, 0, 0,\dots) ,\\
& X_{(m,n,\dots,p,\dots ,l)} = \xi (0,(x^1)^m (x^2)^n\dots  (x^a)^p\dots  (x^{D-2})^l , 0, 0, 0,\dots) ,\\
& Y_{m,n,\dots,p,\dots,l}^1 = \xi (0,0,-(x^1)^{m+1}(x^2)^n\dots(x^a)^p\dots(x^{D-2})^l,0,0,0,\dots) ,\\
& Y_{m,n,\dots,p,\dots,l}^2 = \xi (0,0,0,-(x^1)^m(x^2)^{n+1}\dots(x^a)^p\dots(x^{D-2})^l,0,0,\dots) ,\\
& Y_{m,n,\dots,p,\dots,l}^a = \xi (0,0,0,0,\dots,-(x^1)^m(x^2)^n\dots(x^a)^{p+1}\dots(x^{D-2})^l,0,\dots) ,\\
& Y_{m,n,\dots,p,\dots,l}^{D-2} = \xi (0,0,0,\dots-(x^1)^m(x^2)^n\dots(x^a)^p\dots(x^{D-2})^{l+1}) .
\end{split}
\end{equation}
The nonvanishing commutation relations read
\begin{align}
\label{cca}
&[Y_{m,n,\dots,p,\dots,l}^a,Y_{m',n',\dots,p',\dots,l'}^a] = (p-p')Y_{m+m',n+n',\dots,p+p',\dots,l+l'}^a , \notag\\
&[Y_{m,n,\dots,p,\dots,k,\dots,l}^a,Y_{m',n',\dots,p',\dots,k',\dots,l'}^b] =k Y_{m+m',n+n',\dots,p+p',\dots,k+k'\dots,l+l'}^a-p' Y_{m+m',n+n',\dots,p+p',\dots,l+l'}^b \notag,\\
&[Y_{m,n,\dots,p,\dots,l}^a,T_{m',n',\dots,p',\dots,l'}] =-p T_{m+m',n+n',\dots,p+p',\dots,l+l'}, \notag\\
&[Y_{m,n,\dots,p,\dots,l}^a,X_{m',n',\dots,p',\dots,l'}] =-p X_{m+m',n+n',\dots,p+p',\dots,l+l'}, \notag\\
&[X_{m,n,\dots,p,\dots,l},T_{m',n',\dots,p',\dots,l'}] =\kappa X_{m+m',n+n',\dots,p+p',\dots,l+l'}. \notag
\end{align}
Let \(Y^a_\textbf{m}\equiv Y^a_{m_1,m_2,\dots,m_a,\dots,m_{D-2}}, T_\textbf{m}\equiv T_{m_1,m_2,\dots,m_a,\dots,m_{D-2}}\) and \(X_\textbf{m}\equiv X_{m_1,m_2,\dots,m_a,\dots,m_{D-2}}\) with \(\textbf{m}=(m_1,m_2,\dots,m_a,\dots,m_{D-2})\) standing for the list of indices, one can cast the above algebra into the following form
\begin{equation}
\begin{split}
\label{gwa}
&[Y^i_\textbf{m},Y^j_\textbf{n}]=m_j Y^i_{\textbf{m}+\textbf{n}}-n_i Y^j_{\textbf{m}+\textbf{n}} ,\\
&[Y^i_\textbf{m},T_\textbf{n}]=-n_i T_{\textbf{m}+\textbf{n}} ,\\
&[Y^i_\textbf{m},X_\textbf{n}]=-n_i X_{\textbf{m}+\textbf{n}} ,\\
&[T_\textbf{m},X_\textbf{n}]=-\kappa X_{\textbf{m}+\textbf{n}} .
\end{split}
\end{equation}
The first line of the algebra represents a generalization of the Witt algebra. Its presence at the black hole horizons was only recently noticed in a remarkably different construction \cite{JWM3}.

As a side remark, (\ref{gwa}) has a nice subalgebra,
\begin{equation}
\begin{split}
\label{sagb}
&[\wtd{Y}^i_\textbf{m},\wtd{Y}^j_\textbf{n}]=(m_j-n_i) \wtd{Y}^i_{\textbf{m}+\textbf{n}}\delta_{i,j} ,\\
&[\wtd{Y}^i_\textbf{m},T_\textbf{n}]=-n_i T_{\textbf{m}+\textbf{n}} ,\\
&[\wtd{Y}^i_\textbf{m},X_\textbf{n}]=-n_i X_{\textbf{m}+\textbf{n}} ,\\
&[T_\textbf{m},X_\textbf{n}]=-\kappa X_{\textbf{m}+\textbf{n}} ,
\end{split}
\end{equation}
where $\wtd{Y}^i_\textbf{m}\equiv  Y^a_{0,\dots,m_a,\dots,0}$.

The generators \(T\) and \(X\) are two copies of supertranslation currents \cite{IPBH} associated with the symmetry
\begin{equation}
\upsilon\rightarrow \upsilon + T(x^A) + e^{-\kappa \upsilon} X(x^A) ,
\end{equation}
and the vector fields \(Y^A\) are responsible for generating generalized superrotations
\begin{equation}
x^A\rightarrow x^A + Y^A(x^E) .
\end{equation}
Note that there are two sets of supertranslation currents given by \(X_\textbf{m}\) and  \(T_\textbf{m}\), and $D-2$ sets of generalized superrotations \(Y_\textbf{m}^i\) which generate a new algebra that can be regarded as a type of generalization of the usual Witt algebra. This general extension of Witt algebra also appears in the discussion about internal gauge symmetry in higher dimension\cite{JWM3}. From the last line of Eq.(\ref{gwa}),  \(X_\textbf{m}\) can be viewed as an expansion under the action of \(T_\textbf{0}\).

It is easy to check that the subalgebra (\ref{sagb}) can be viewed as a direct product of the algebra found in \cite{ESBH}  with more generalized superrotations. In the four dimensional case, the metric of \(D-2\) hypersurface \(\Omega_{AB}\) can be written by stereographic coordinates \(x^a=(z,\bar{z})\), in a such way that
\begin{equation}
\label{sc}
\Omega_{AB}=\frac{4\Omega}{(1+z\bar{z})^2}dz d\bar{z}
\end{equation}
This implies that \(Y^A\) are conformal Killing vectors on the \(\Omega_{AB}\) in three and four dimensional case, then we can always choose the form of \(Y^A=Y^A(x^A)\), i.e.\(Y=Y(z),\bar{Y}=\bar{Y}(\bar{z})\), to agree with the conditions $\wtd{Y}^i_\textbf{m}\equiv  Y^a_{0,\dots,m_a,\dots,0}$ automatically. Unfortunately, this calculation is not suitable for \(D\) dimensional cases, as there does not exist a special form similar with (\ref{sc}) for \(\Omega_{AB}\), and \(Y^A\)  are no longer conformal Killing vectors on it.

\section{THE CASE OF STATIONARY BLACK HOLES}
In this section, we apply the above results to the case of stationary black holes.

\subsection{Near horizon metrics for D-dimensional stationary black holes}
Consider the general metric for a stationary and axisymmetric black hole \cite{JWM2,JWM3,JWM1,0811.2225},
\begin{align}
\label{gsam}
ds^2 = f(-\dfrac{\Delta}{V^2}dt^2+\dfrac{1}{\Delta}dr^2)+h_{ij}d\theta^i d\theta^j + g_{ab}(d\phi^a - \omega^a dt)(d\phi^b - \omega^b dt)
\end{align}
where \(f,V,h_{ij},g_{ab}\) and \(\omega^a\) are functions of \(r\) and \(\theta^i\). \(\Delta\) only depends on \(r\) and the horizon located at \(\Delta(r_0)=0\). In principle, one can identify the coordinates as the asymptotic time \(t\), the radial coordinate \(r\), the latitudinal angles \(\theta^i(i=1,\dots,[\frac{D}{2}]-1)\) and the azimuthal angles \(\phi^a(a=1,\dots,[\frac{D+1}{2}]-1)\), where $D$ is the total dimension of the spacetime. In the near horizon region, \(V\) and \(\omega^a\) can always be expanded in the form of:
\begin{align}
&V(r,\theta^i)=V_0(r_0)+V_1(\theta^i)\Delta+\mathcal{O}(\Delta^2)\\
&\omega^a(r,\theta^i)=\omega^a_0(r_0)+\omega^a_1(\theta^i)\Delta+\mathcal{O}(\Delta^2)
\end{align}

The task of this subsection is to find out the connection between Eq.(\ref{metric1}) and Eq.(\ref{gsam}). Appendix A of \cite{SNIDTH} demonstrates the construction of the Gaussian null coordinate system for a Kerr-Newman horizon. By extending the result of \cite{SNIDTH} to arbitrary dimensional stationary black hole horizon, the functions in Eq.(\ref{metric1}) can be expressed by the functions in Eq.(\ref{gsam}). A pair of future-oriented null vector \(l^a\) and \(n^a\) with the condition of \(l^a=\frac{\partial}{\partial \upsilon}\) and \(l^a n_a=-1\) on the horizon are necessary to get the metric under the construction of \cite{SNIDTH}.

Using the coordinate transformation
\begin{align}
&\upsilon=t+\int^r \dfrac{V_0(r')}{\Delta(r')}dr'\notag\\
&\varphi^a=\phi^a+\int^r \dfrac{V_0(r')\omega^a_0(r')}{\Delta(r')}dr'-\omega^a_0(r_0)\upsilon
\end{align}
the metric(\ref{gsam}) is transformed into:
\begin{align}
\label{mfsc}
ds^2&=(-f\dfrac{\Delta}{V^2}+g_{ab}(\omega^a-\omega^a_0(r_0)(\omega^b-\omega^b_0(r_0)))d\upsilon^2+2(\frac{fV_0}{V^2}-\frac{V_0}{\Delta}g_{ab}(\omega^a-\omega^a_0)(\omega^b-\omega^b_0(r_0))d\upsilon dr \notag\\
&+(\dfrac{V^2-V_0^2}{\Delta V^2}f+\dfrac{V_0 ^2}{\Delta^2}g_{ab}(\omega^a-\omega^a_0)(\omega^b-\omega^b_0))dr^2\notag-2g_{ab}(\omega^a-\omega^a_0(r_0))d\varphi^b d\upsilon\\
&+2\frac{V_0}{\Delta}g_{ab}(\omega^a-\omega^a_0)d\varphi^b dr+h_{ij}d\theta^i d\theta^j+g_{ab}d\varphi^a d\varphi^b
\end{align}
The horizon is a null hypersurface with fixed \(r=r_0\), then the \((D-1)\) dimensional metric on \(H\) reads:
\begin{align}
\label{mfh}
dS^2=h_{ij}d\theta^i d\theta^j+g_{ab}d\varphi^a d\varphi^b
\end{align}

It is easy to check that \(\frac{\partial}{\partial \upsilon}\) is one of the null-normal vectors on the horizon. The second null vector is:
\begin{align}
\label{efn}
n=-\dfrac{1}{2}g_{ab}(\theta^i,r_0)\omega_0^a(r_0)\omega_0^b(r_0)\frac{\partial}{\partial \upsilon}-\frac{V_0(r_0)}{f(\theta^i,r_0)}\frac{\partial}{\partial r}-\omega^a_0(r_0)\frac{\partial}{\partial \varphi^a}
\end{align}
Consider a null geodesic congruence which crosses the horizon with tangent vector field $\textbf{n}$, marked by the points on them, and parametrized with affine parameter $\rho$, such that $\rho = 0$ identifies the horizon. Near the horizon, those geodesics are expanded to the second order of $\rho$
\begin{align}
X^\alpha_{(\upsilon,r,\theta,\phi)} \approx X^\alpha\mid_{\rho = 0} + \rho \frac{dX^\alpha}{d\rho}\mid_{\rho = 0} + \frac{\rho^2}{2} \frac{d^2 X^\alpha}{d \rho^2}\mid_{\rho=0}
\end{align}
This defines a coordinate transformation from $(\upsilon_1,\rho,\theta_1^i,\phi_1^a)$ to $(\upsilon,r,\theta^i,\phi^a)$, with
\begin{align}
X^\alpha \mid_{\rho=0} = [\upsilon_1,r_0,\theta_1^i,\phi_1^a] \notag
\end{align}
And $X^\alpha$ satisfied the geodesics equation and the null vector $\textbf{n}$ is tangent to it on the horizon, this implies:
\begin{align}
\frac{d X^\alpha}{d\rho}\mid_{\rho=0}= n^\alpha
\end{align}
and
\begin{align}
\frac{d^2 X^\alpha}{d \rho^2} + \Gamma^\alpha_{\beta \gamma} \frac{dX^\beta}{d\rho} \frac{dX^\gamma}{d\rho} = 0 \Rightarrow \frac{d^2 X^\alpha}{d \rho^2}\mid_{\rho=0}=- \Gamma^\alpha_{\beta \gamma}\mid_{\rho=0} n^\beta n^\gamma
\end{align}
Then the first order expansion of the metric is
\begin{align}
g_{\mu \nu} \approx g^{(0)}_{\mu \nu} +  \rho  g^{(1)}_{\mu \nu}
\end{align}
The zeroth order components are:
\begin{align}
g^{(0)}_{\upsilon_1 \rho} = -1 \notag\\
g^{(0)}_{ij}=h_{ij}    \notag \\
g^{(0)}_{ab}=g_{ab}     \notag
\end{align}
The first order components are
\begin{align}
g_{\upsilon_1 \upsilon_1}^{(1)} &= \frac{1}{V_0} \frac{\partial \Delta}{\partial r}\mid_{r=r_0} \notag \\
g_{\upsilon_1 i}^{(1)} &= \frac{1}{f} \frac{\partial f}{\partial\theta^i}\mid_{r=r_0} \notag\\
g_{\upsilon_1 a}^{(1)} &= g_{ab} \frac{V_0}{f} \frac{\partial \omega^b}{\partial r} \mid_{r=r_0} \notag \\ g_{ij}^{(1)} &= -\frac{V_0}{f} \frac{\partial h_{ij}}{\partial r} \mid_{r=r_0} \notag \\
g_{ia}^{(1)} &= -\frac{V_0(r_0)}{f(r_0)} \frac{V_0}{\Delta} g_{ab}(\omega^b-\omega^b_0)\frac{\partial f(r_0)}{\partial \theta^i} \notag \\
g_{ab}^{(1)} &= -\frac{V_0{(r_0)}}{f(r_0)} \frac{\partial g_{ab}}{\partial r}\mid_{r=r_0} \notag
\end{align}
the components not listed above vanish. With these results, Eq.(\ref{metric1}) can describe the geometry of the near horizon region for arbitrary dimensional stationary black holes.

\subsection{Charge}
The charge Eq.(\ref{vscd}) for the stationary black holes are
\begin{equation}
\cancel{\delta} Q_{[T;Y^A]}=\dfrac{1}{16\pi G}\int(d^{D-2}x)_{\upsilon\rho}2[2T\kappa \delta(\sqrt{\det\Omega})-Y^A\delta(\theta_A\sqrt{\det\Omega})]
\end{equation}
As \(\delta T=0,\delta Y^A=0, \delta\kappa=0\), the charges can be integrated:
\begin{equation}
\label{cfs}
Q_{[T;Y^A]}=\dfrac{1}{16\pi G}\int(d^{D-2}x)_{\upsilon\rho}2\sqrt{-\bar{g}}[2T\kappa-Y^A\theta_A] + Q_0
\end{equation}
Those changes close under Poisson bracket
\begin{equation}
\begin{split}
\label{pb}
\{Q(T_1,Y^A_1),Q(T_2,Y^A_2)\}=Q(T_{12},Y^A_{12})
\end{split}
\end{equation}

Defining
\begin{equation}
\begin{split}
\label{cem}
& \mathcal T_\textbf{m} = Q ((x^1)^{m_1} (x^2)^{m_2} \dots (x^a)^{m_a} \dots (x^{D-2})^{m_{D-2}}  , 0, 0,0, 0,\dots)\\
& \mathcal Y_\textbf{m}^a = Q (0,0,0,\dots ,-(x^1)^{m_1} (x^2)^{m_2} \dots (x^a)^{m_a+1} \dots (x^{D-2})^{m_{D-2}},0,\dots)\\
\end{split}
\end{equation}
And define the generator
\begin{equation}
\mathcal P_{(p,q,\dots,k,\dots,l)}=\sum\limits_{m\in Z}\sum\limits_{n\in Z}\dots \sum\limits_{o\in Z}\dots \sum\limits_{s\in Z} \mathcal T_{(m,n,\dots,o,\dots ,s)}\mathcal T_{(p-m,q-n,\dots,k-o,\dots ,l-s)},
\end{equation}
and setting \(Q_0=0\), one can obtain:
\begin{equation}
\begin{split}
\label{cafs}
&[Y^i_\textbf{m},Y^j_\textbf{n}]=m_j Y^i_{\textbf{m}+\textbf{n}}-n_i Y^j_{\textbf{m}+\textbf{n}} \\
&[Y^i_\textbf{m},P_\textbf{n}]=(m_i-n_i) P_{\textbf{m}+\textbf{n}} \\
\end{split}
\end{equation}
Notice that, in four dimensional cases, this algebra reduce to extended BMS algebra found in \cite{ESBH,STSR}.

\subsubsection{D=2n+2}
In this case, the number of latitudinal angle \(\theta^i\) equal to the number of azimuthal angle \(\phi^a\),  i.e. \(D_{\theta^i}=D_{\phi^a}=n\). Label both kinds of coordinates with \(a,b,c,\dots\). To calculate the charges which are defined in term of the coordinate \(x^a\), we let:
\begin{align}
\label{ctf}
x^a=e^{i\phi^a}u_a(\theta^a); \bar{x}^a=e^{-i\phi^a}u_a(\theta^a)
\end{align}
Then:
\begin{align}
\label{rbct}
A=\int_{r=r_0}(d^{D-2}x)_{\upsilon\rho}2\sqrt{-\bar{g}}=\int_{r=r_0}(d^{D-2}x)_{tr}2 \sqrt{hg}
\end{align}

The difference between the integration in phase space of two black holes at fixed \(\kappa\) is
\begin{align}
\label{eft}
\mathcal T_\textbf{0}^{(A)}-\mathcal T_\textbf{0}^{(B)}=\dfrac{2\kappa}{16\pi G}\int_{r=r_0}(d^{D-2}x)_{\upsilon\rho}2 \delta\sqrt{-\bar{g}}=\dfrac{\kappa}{2\pi}\dfrac{\Delta A}{4G}=T\Delta S_{BH}
\end{align}
Notice that for a stationary black hole, Hawking temperature:
\begin{align}
T=\frac{1}{4\pi}\frac{1}{V}\frac{\partial \Delta}{\partial r}|_{r=r_0}=\dfrac{\kappa}{2\pi}\notag
\end{align}
This means that the zero mode of supertranslation charge is intimately related to the Bekenstein-Hawking entropy and Hawking temperature. It should be noticed that \(Q_0=0\) is not necessary in this calculation.

For the charges associate with generalized superrotations. Note :
\begin{align}
\theta_A dx^A=\theta_{x^a} dx^a + \theta_{x^{\bar{a}}}dx^{\bar{a}}=\theta_{\theta^a}d\theta^a+\theta_{\phi^a}d\phi^a
\end{align}
which leads to:
\begin{align}
x^a\theta_{x^a}=\frac{1}{2}(\frac{u_a(\theta^a)}{u_a'(\theta^a)}\theta_{\theta^a}-i\theta_{\phi^a})\notag\\
\bar{x}^a\theta_{\bar{x}^a}=\frac{1}{2}(\frac{u_a(\theta^a)}{u_a'(\theta^a)}\theta_{\theta^a}+i\theta_{\phi^a})\notag
\end{align}
From Eq.(\ref{cfs}), the charges associate with generalized superrotation can be written as:
\begin{eqnarray}
\label{SRI}
\mathcal{Y}^a_{(m_1,m_1',\dots,m_a,m_a',\dots)}&=&\frac{1}{16\pi G}\int_{r-r_0}(d^{D-2}x)_{\upsilon\rho}\sqrt{-\bar{g}}(\frac{u_a(\theta^a)}{u_a'(\theta^a)}\theta_{\theta^a}-i\theta_{\phi^a})\nn\\
&&u^{2m_1}_1(\theta^1)\dots u^{2m_a}_a(\theta^a)\dots \delta_{m_1,m_1'}\dots\delta_{m_a,m_a'}\dots\\
\mathcal{\bar{Y}}_{(m_1,m_1',\dots,m_a,m_a',\dots)}^a&=&\frac{1}{16\pi G}\int_{r-r_0}(d^{D-2}x)_{\upsilon\rho}\sqrt{-\bar{g}}(\frac{u_a(\theta^a)}{u_a'(\theta^a)}\theta_{\theta^a}+i\theta_{\phi^a})\nn\\
&&u^{2m_1}_1(\theta^1)\dots u^{2m_a}_a(\theta^a)\dots \delta_{m_1,m_1'}\dots\delta_{m_a,m_a'}\dots
\end{eqnarray}
It is easy to check that:
\begin{align}
Q_{(0,\partial_{\phi^a})}=&i(\mathcal{Y}_\textbf{0}^a-\mathcal{\bar{Y}}_\textbf{0}^a)=-\frac{1}{16\pi G}\int_{r-r_0}(d^{D-2}x)_{\upsilon\rho}2\sqrt{-\bar{g}}\theta_{\phi^a}\notag\\
=&-\frac{1}{16\pi G}\int_{r-r_0}(d^{D-2}x)_{\upsilon\rho}2\sqrt{-\bar{g}}\frac{V}{f}g_{ab}\frac{\partial \omega^b}{\partial r}=J_a
\end{align}
Each pair of \(Y^a_\textbf{0}\) and \(\bar{Y}_\textbf{0}^a\) correspond to a angular momenta. And the nonzero modes of supertranslation are found to be
\begin{equation}
\begin{split}
\label{STI}
\mathcal T_\textbf{m} =\frac{2\kappa}{16\pi G}\int_{r-r_0} &\sqrt{hg}  d\theta^{m_1}\wedge\dots\wedge d\theta^{m_a}\wedge\dots\wedge d\theta^{m_n}\\
&u^{2m_1}_1(\theta^1)\dots u^{2m_a}_a(\theta^a)\dots \delta_{m_1,m_1'}\dots\delta_{m_a,m_a'}\dots
\end{split}
\end{equation}
The integrations \ref{SRI} to be vanish and \ref{STI} to be divergent for nonzero-modes in four dimensional cases, see \cite{STSR,BMSSA,CHCA}.

\subsubsection{D=2n+3}
In this case the number of latitudinal angle \(D_{\theta^a}=n\), and the number of azimuthal angle \(D_{\phi^a}=n+1\). Similar to the calculation above, n pair of \(\theta^a\) and \(\phi^a (a=1,\dots,n)\) can be picked out to define \(Y^a_m\) and \(\bar{Y}^a_m\), leaving a residual coordinate which is named as \(\phi^{n+1}\), producing one more angular momenta:
\begin{eqnarray}
\mathcal{Y}^a_{(m_1,m_1',\dots,m_a,m_a',\dots)}&=&\frac{1}{16\pi G}\int_{r-r_0}(d^{D-2}x)_{\upsilon\rho}\sqrt{-\bar{g}}(\frac{u_a(\theta^a)}{u_a'(\theta^a)} \theta_{\theta^a}-i\theta_{\phi^a})\nn\\
&&u^{2m_1}_1(\theta^1)\dots u^{2m_a}_a(\theta^a)\dots \delta_{m_1,m_1'}\dots\delta_{m_a,m_a'}\dots\delta_{m_{n+1},0}\notag
\end{eqnarray}
Similar with the \(D=2n+2\) cases,
\begin{align}
& Q_{(0,\partial_{\phi^{n+1}})}=i\mathcal{Y}_0^{n+1}=\frac{-1}{16\pi G}\int_{r-r_0}(d^{D-2}x)_{\upsilon\rho}2\sqrt{-\bar{g}}\theta_{\phi^{n+1}}
=J_{n+1} .\notag
\end{align}

In the case of D-dimensional stationary black holes, zero-modes are related to entropy and angular momenta.
\section{SUMMARY}
In this paper, we extend the work of \cite{ESBH} to the case of black holes in generic dimensions. We start by considering the boundary conditions for arbitrary dimensional black holes proposed in \cite{ESBH,STSR}. We calculate asymptotic Killing vectors, forming a closed algebra by using a modified version of Lie brackets mentioned in \cite{ABCC}. Then we calculate the surface charges and asymptotic symmetry group, which contains two supertranslations and $D-2$ generalized superrotations. Remarkably, the superrotations generate a generalized Witt algebra that was previously found on the black hole horizons in a very different construction \cite{JWM3}, indicating that there might be some connection between the two different ways of treatment. We have applied the result to the case of $D$-dimensional stationary black holes by calculating the charge algebra and zero-modes.

\section*{ACKNOWLEDGMENTS}
The authors thank Xun Wang and Jiandong Zhang for helpful discussions. This work was supported by the National Natural Science Foundation of China (Grant No. 11475064).

\end{document}